\documentclass[12pt]{article}
\usepackage{epsfig,a4}

 \advance\oddsidemargin by -1in
 \advance\evensidemargin
 by -1in
 \marginparsep
 0.4cm
 \advance\topmargin by
 -0.0in
 \makeindex

 \newcommand\la{\langle}
 \newcommand\ra{\rangle}
 \newcommand\beq{\begin{equation}}
 
 \newcommand\eeq{\end{equation}}                                               
 \newcommand\beqn{\begin{eqnarray}}
 \newcommand\eeqn{\end{eqnarray}}
 \newcommand\GeV{{\rm GeV}}

\def\BA{\begin{eqnarray}}
\def\BE{\begin{equation}}
\def\BF{\begin{figure}[htb]}
\def\BT{\begin{table}[htb]}
\def\EA{\end{eqnarray}}
\def\EE{\end{equation}}
\def\EF{\end{figure}}
\def\ET{\end{table}}

\def\la{\langle}
\def\ra{\rangle}
\def\mb{\,\mbox{mb}}
\def\fm{\,\mbox{fm}}
\def\GeV{\,\mbox{GeV}}

\def\lsim{\mathrel{\rlap{\lower4pt\hbox{\hskip1pt$\sim$}}
    \raise1pt\hbox{$<$}}}         
\def\gsim{\mathrel{\rlap{\lower4pt\hbox{\hskip1pt$\sim$}}
    \raise1pt\hbox{$>$}}}         

\begin{document} 
\vspace*{3cm}

\date{today}

\begin{center}
{\LARGE\bf 
Incoherent Production of Charmonia off Nuclei 
\vspace*{0.2cm}
As a Good Tool \\
\vspace*{0.2cm}
for Study of Color Transparency}
\end{center}


\begin{center}

\vspace{0.5cm}
 {\large J.~Nemchik}
 \\[1cm]
 {\sl Institute of Experimental Physics SAS, Watsonova 47,
04353 Kosice, Slovakia}

\end{center}

\vspace{1cm}

\begin{abstract}

Within a light-cone QCD formalism which 
incorporates color transparency (CT), 
coherence length (CL) effects 
and gluon shadowing (GS) we study electroproduction
of charmonia off nuclei.
In contrast to production of light vector mesons 
($\rho^0$, $\Phi^0$) when 
at small and medium energies CT and the onset of CL effects are not
easily separated, in production of charmonia CT effects dominate.
We found rather large CT effects
in the range of $Q^2 \leq 20\GeV^2$.
They are stronger at low than at high energies 
and can be easily identified by the planned future
experiments.
Our parameter-free calculations
explain well the NMC data for variation 
with photon energy of the $Sn/C$ ratio
of nuclear transparencies.
We provide 
predictions for incoherent and coherent 
production of charmonia for future
measurements.

\end{abstract}


\newpage



\section{Introduction: space-time pattern
of charmonium production} 
\label{intro}

The dynamics of charmonium production has been 
a hot topic evolved
intensively during almost three last decades.
Discovery of $J/\Psi$ in 1973 confirmed the idea
of charm quark and gave a basis for its further
investigations, which were
affected also by new experiments situated
on more developed accelerators using more
and more powerful electronics.
Later, at the beginning of 90's  
the experiments with
relativistic heavy-ion collisions \cite{heavyion}
stimulated the enhanced interest about 
charmonium suppression as a possible
indication of the quark-gluon plasma formation.
That fact has given a further motivation to continue in
investigation of space-time pattern of
charmonium production and opened new possibilities to analyze 
various corresponding phenomena.

One of the fundamental phenomenon coming from QCD is 
color transparency (CT) studied intensively almost two last decades.
This phenomenon can be treated
either in the hadronic or in the quark basis.
The former approach leads to
Gribov's inelastic corrections \cite{gribov},
the latter one manifests itself as a result of 
color screening \cite{zkl,bbgg}. 
Although these two approaches are
complementary, the quark-gluon interpretation is more intuitive and
straightforward. 
Colorless hadrons can interact only because color
is distributed inside them. 
If the hadron transverse size $r$ tends to zero then
interaction cross section $\sigma(r)$ vanishes as $r^2$
\cite{zkl}.
As a result the nuclear medium is more transparent 
for smaller transverse size of the hadron.
Besides, this fact naturally
explains the correlation between the cross sections of hadrons and their
sizes \cite{gs,hp,p}. 

Investigation of diffractive electroproduction of vector mesons 
off nuclei is very effective and sensitive  for study of CT. 
A photon of high virtuality $Q^2$ is
expected to produce a pair with a small $\sim 1/Q^2$ transverse
separation\footnote{For production of light vector mesons
($\rho^0$, $\Phi^0$) very asymmetric pairs can be possible 
when the $q$ or $\bar q$ carry almost the whole
photon momentum.
As a results the $\bar qq$ pair can have a large separation,
see Sect.~\ref{lc} and Eq.~(\ref{212}).
Not so for production of charmonia where one can use
the nonrelativistic approximation, $\alpha = 0.5$,
with rather high accuracy.}. 
Then CT manifests itself as 
a vanishing absorption of the
small sized colorless $\bar qq$ wave
packet during propagation through the nucleus. 
Dynamical evolution of the small size $\bar qq$ pair to a normal
size vector meson is controlled by the time scale called formation
time.
Due to uncertainty principle, one needs time interval to resolve
different levels $V$ (the ground state) or $V'$ 
(the next excited state) in the final state.
In the rest frame of the nucleus this formation time is
Lorentz dilated,
 \beq
t_f = \frac{2\,\nu}
{\left.m_V^\prime\right.^2 - m_V^2}\ ,
\label{20}
 \eeq
where $\nu$ is the photon energy.
A rigorous quantum-mechanical description of the pair evolution was
suggested in \cite{kz-91} and is based on the light-cone Green function
technique. 
A complementary description of the same process in the hadronic basis
is presented in \cite{hk-97}.

Another phenomenon known to cause nuclear
suppression is the effect of quantum coherence.
It results 
from destructive interference of
the amplitudes for which the interaction takes place on different bound
nucleons. It reflects the distance from the absorption 
point
when the pointlike photon becomes the hadronlike $\bar qq$ pair.
This may be also interpreted as a 
lifetime of $\bar qq$ fluctuation providing the time scale which 
controls shadowing.
Again, it can be estimated relying on the uncertainty principle and
Lorentz time dilation as,
 \beq
t_c = \frac{2\,\nu}{Q^2 + m_V^2}\ .
\label{30}
 \eeq 
 It is usually called coherence time, but we also will use the term
coherence length (CL), since light-cone kinematics is assumed, $l_c=t_c$
(similarly, for formation length $l_f=t_f$). CL is related to the
longitudinal momentum transfer $q_c=1/l_c$ in $\gamma^*\,N \to V\,N$,
which controls the interference of the production amplitudes from
different nucleons.

Exclusive production of vector mesons at high energies is controlled by
the small-$x_{Bj}$ physics, and gluon shadowing becomes an important
phenomenon \cite{knst-01}. 
It was shown in \cite{ikth-02} that for electroproduction of charmonia off 
nuclei 
the gluon shadowing starts to be important at c.m.s. energy
$\sqrt{s} \geq 30-60\GeV$ in dependence on nuclear target and $Q^2$.
Although the gluon shadowing within the kinematic range 
important for investigation of CT and discussed in the 
present paper is quite small we include it in all the calculations.

In electroproduction of
vector mesons off nuclei one needs to disentangle CT (absorption) 
and CL (shadowing) as the two sources of nuclear suppression. 
Detailed analysis of the CT and CL effects in electroproduction
of vector mesons off nuclei showed \cite{knst-01}
that one can easily identify the difference of the nuclear suppression
corresponding to absorption and shadowing in the
two limiting cases for the example of vector dominance model (VDM).\\
i.) The limit of $l_c$, shorter than the mean
internucleon spacing $\sim 2\fm$. In this case only final state
absorption matters. The ratio of the quasielastic (or incoherent)
$\gamma^*\, A \to V\,X$ and $\gamma^*\, N \to V\,X$ cross sections,
usually called nuclear transparency, reads \cite{kz-91},
 \beqn
Tr_A^{inc}\Bigr|_{l_c\ll R_A} &\equiv& 
\frac{\sigma_V^{\gamma^*A}}
{A\,\sigma_V^{\gamma^*N}} =
\frac{1}{A}
\,\int d^2b\,
\int\limits_{-\infty}^{\infty}
dz\,\rho_A(b,z)\,
\exp\left[-\sigma^{VN}_{in}
\int\limits_z^{\infty} dz'\,
\rho_A(b,z')\right]\nonumber\\
&=& \frac{1}{A\,\sigma^{VN}_{in}}\,
\int d^2b\,\left\{1 - 
\exp\Bigl[-\sigma^{VN}_{in}\,T(b)\Bigr]\right\}=
\frac{\sigma^{VA}_{in}}{A\,\sigma^{VN}_{in}}\ .
\label{40}
 \eeqn
Here $z$ is the longitudinal coordinate and $\vec{b}$ is the impact
parameter of the production point of vector meson. 
In (\ref{40}) $\rho_A(b,z)$ is the 
nuclear density and $\sigma_{in}^{VN}$ is   
the inelastic $VN$ cross section.

ii.) 
In the limit of long $l_c$ expression for
nuclear transparency takes a different form,
 \beq
Tr_A^{inc}\Bigr|_{l_c\gg R_A} = 
\int d^2b\,T_A(b)\,
\exp\left[-\sigma^{VN}_{in}\,T_A(b)\right]\ ,
\label{50}
 \eeq 
where we assume $\sigma^{VN}_{el} \ll \sigma^{VN}_{in}$ for the
sake of simplicity.  $T_A(b)$ is the nuclear thickness function
 \beq
T_A(b) = \int\limits_{-\infty}^{\infty} dz\,\rho_A(b,z)\ .
\label{60}
 \eeq 
The exact expression which interpolates between the two
regimes (\ref{40}) and (\ref{50}) can be found in \cite{hkn}.

The problem of separation of CT and CL effects
arises especially in production of light vector
mesons ($\rho^0$, $\Phi^0$) \cite{knst-01}.
In this case the coherence length is larger or
comparable with the formation one 
starting from the photoproduction limit up to
$Q^2 \sim 1\div 2\GeV^2$.
In charmonium production, however, there is a strong
inequality $l_f > l_c$
independently of $Q^2$ and $\nu$. It leads to
a different scenario of CT-CL mixing as compared
with production of light vector mesons. 
That fact gives a motivation for separated study of
$J/\Psi$ production presented in this paper using
light-cone dipole approach generalized for the case
of a finite coherence length and developed in \cite{knst-01}.
Another reason is supported by the recent paper 
\cite{ikth-02} where charmonium production was calculated
in the approximation of long coherence length $l_c \gg R_A$
using realistic charmonia wave functions and
corrections for finite values of $l_c$.
It gives very interesting possibility to compare
the predictions of the present paper with the results
obtained from \cite{ikth-02}
for enhancement of reliability of theoretical predictions
as a realistic basis for planned future
experiments of electron-nucleus collisions.

The paper is organized as follows.
In Sect.~\ref{lc} we present a short review of the light-cone (LC) 
approach to diffractive
electroproduction of vector mesons in the rest frame of the nucleon
target. Here we also present the individual ingredients contained
in the production amplitude: (i) the dipole cross section
characterizing the universal interaction 
cross section for a colorless quark-antiquark dipole and a 
nucleon. (ii)  
The LC wave function for a quark-antiquark fluctuation of the virtual
photon. (iii) The LC wave function of charmonia.

As the first test of the model we calculate in Sect.~\ref{data-N} the
cross section of elastic electroproduction of $J/\Psi$ 
off a nucleon target.  These parameter-free calculations reproduce both
energy and $Q^2$ dependence remarkably well, including the absolute
normalization. 

Sect.~\ref{psi-incoh} is devoted to incoherent production of $J/\Psi$
off nuclei. 
Numerical calculations 
are compared with the available NMC data for variation with photon energy
of the $Sn/C$ ratio of nuclear transparencies. 
We find a different scenario of an interplay
between coherence and formation length effects
from one occurred in production of light vector mesons.
Because a variation of
$l_c$ with $Q^2$ can mimic CT at medium and low energies, one can map
experimental events in $Q^2$ and $\nu$ in such a way as to keep
$l_c=const$. 
The LC dipole formalism 
predicts rather large effect of CT in the range of $Q^2 \leq 20\GeV^2$. 
This fact makes
it feasible to find a clear signal of CT effects also in exclusive 
production of $J/\Psi$ in the planned future experiments.

Coherent production of vector mesons off nuclei leaving the nucleus
intact is studied in Sect.~\ref{psi-coh}. 
The detailed calculations show that
the effect of CT on the $Q^2$
dependence of nuclear transparency at $l_c=const$ is weaker than in the
case of incoherent production and is difficult to be detected at low
energies since the cross section is small. 

Although the gluon shadowing starts to manifest itself at $\sqrt{s}\geq 
30-60\GeV$ and is not very significant in the energy range
important for searching for CT effects
in the present paper we include it in all calculations.

The results of the paper are summarized and discussed in
Sect.~\ref{conclusions}. An optimistic prognosis for discovery of CT 
in electroproduction of charmonia is made for 
the future experiments.

\section{A short review of the light-cone dipole phenomenology for elastic 
electroproduction of charmonia \boldmath$\gamma^{*}N\to J/\Psi~N$}
\label{lc}

The light-cone (LC) dipole approach for elastic electroproduction 
$\gamma^{*}N\to V~N$ was already used in papers
\cite{hikt-00} for study of exclusive photo- and electroproduction
of charmonia and  
in \cite{knst-01} for elastic virtual photoproduction of
light vector mesons $\rho^0$ and $\Phi^0$.
Therefore, we present only a short review of this LC phenomenology
with the main emphasis to elastic electroproduction of charmonia.
Here a diffractive process
is treated as elastic scattering of a $\bar qq$ fluctuation
($\bar cc$ fluctuation for the case of charmonium production) of the
incident particle.  The elastic amplitude is given by convolution of the
universal flavor independent dipole cross section for the $\bar qq$
interaction with a nucleon, $\sigma_{\bar qq}$, \cite{zkl}
and the initial and final wave functions.  
For the exclusive photo- or
electroproduction of charmonia $\gamma^{*}N\to J/\Psi~N$ 
the forward production amplitude is
represented in the quantum-mechanical form
 \BE
{\cal M}_{\gamma^{*}N\rightarrow J/\Psi~N}(s,Q^{2}) =
\langle J/\Psi |\sigma_{\bar qq}^N(\vec r,s)|\gamma^{*}\rangle=
\int\limits_{0}^{1} d\alpha \int d^{2}{{r}}\,\,
\Psi_{J/\Psi}^{*}({\vec{r}},\alpha)\,   
\sigma_{\bar qq}({\vec{r}},s)\,  
\Psi_{\gamma^{*}}({\vec{r}},\alpha,Q^2)\,
\label{120}
 \EE
 with the normalization 
 \beq
\left.\frac{d\sigma}{dt}\right|_{t=0} =
\frac{|{\cal M}|^{2}}{16\,\pi}.
\label{125}
 \eeq

In order to calculate the photoproduction amplitude one needs to know the
following ingredients of Eq.~(\ref{120}): (i)  the dipole cross section
$\sigma_{\bar qq}({\vec{r}},s)$ which depends on the $\bar qq$
transverse separation $\vec{r}$ and the c.m. energy squared $s$. (ii)  
The light-cone (LC)  wave function of the photon
$\Psi_{\gamma^{*}}({\vec{r}},\alpha,Q^2)$ which also depends on the
photon virtuality $Q^2$ and the relative share $\alpha$ of the photon
momentum carried by the quark. (iii) The LC wave function
$\Psi_{J/\Psi}(\vec r,\alpha)$ of $J/\Psi$. 

Note that in the LC formalism the photon and meson wave functions contain
also higher Fock states $|\bar qq\ra$, $|\bar qqG\ra$, $|\bar qq2G\ra$,
etc. 
The effects of higher Fock states are implicitly incorporated
into the energy dependence of the dipole cross section
$\sigma_{\bar qq}(\vec r,s)$ as is given in Eq.~(\ref{120}).

The dipole cross section $\sigma_{\bar qq}(\vec r,s)$ 
represents the interaction of a
$\bar qq$ dipole of transverse separation $\vec r$ with a nucleon
\cite{zkl}.
It is a flavor independent universal function of
$\vec{r}$ and energy and allows to describe in a uniform way various
high energy processes. 
It is known to vanish quadratically
$\sigma_{\bar qq}(r,s)\propto r^2$ as $r\rightarrow 0$ due to color
screening (CT property) and cannot be predicted
reliably because of poorly known higher order pQCD corrections and
nonperturbative effects. 
Detailed discussion about the dipole cross section 
$\sigma_{\bar qq}(\vec r,s)$ 
with emphasis to production of light vector mesons
is presented in \cite{knst-01}.
In electroproduction of charmonia the corresponding transverse
separations of $\bar cc$-dipole reach the values $\leq 0.4\fm$
(semiperturbative region). It means that nonperturbative effects
are sufficiently smaller as compared with light vector mesons.
Similarly, the relativistic correstions are also small enough to use safely
the nonrelativistic limit $\alpha = 0.5$ with rather high accuracy
\cite{kz-91}.

There are two popular parameterizations of $\sigma_{\bar qq}(\vec r,s)$.
The first one suggested in \cite{gbw} reflects the fact that
at small separations the dipole cross section should be a function of $r$
and $x_{Bj}\sim 1/(r^2s)$ to reproduce Bjorken scaling. 
It well describes data for DIS at small $x$ and medium and high $Q^2$.
However, at small $Q^2$ it cannot be correct since it predicts energy
independent hadronic cross sections. Besides, $x_{Bj}$ is not any more a
proper variable at small $Q^2$ and should be replaced by energy.
This defect is removed by
the second parameterization suggested in \cite{kst2},
which is similar to one in \cite{gbw}, but contains an explicit
dependence on energy. It is valid down to the limit of real
photoproduction.
Since we want to study CT effects starting from $Q^2 = 0$,
we choose the second parametrization, which has the following form :
 \BA
  \sigma_{\bar qq}(r,s) &=& \sigma_0(s) 
\left[1 - e^{- r^2/r_{0}^2(s)}
  \right]\ ,
  \label{130}
 \EA
where
 \BE
  \sigma_0(s) = \sigma^{\pi p}_{tot}(s)
  \left[1+\frac38 \frac{r_{0}^2(s)}
{\left<r^2_{ch}\right>}\right]\mb\ \\
  \label{140}
 \EE
and
 \BE
  r_0(s)   = 0.88 \left(\frac{s}{s_0}\right)^{\!\!-0.14}  \fm\ .
  \label{150}
 \EE
 Here $\left<r^2_{ch}\right>=0.44\fm^2$ is the mean pion charge radius
squared; $s_0 = 1000\GeV^2$.  The cross section $\sigma^{\pi p}_{tot}(s)$
was fitted to data in \cite{dl,pdt},
 \BE
\sigma^{\pi p}_{tot}(s) = 
23.6\,\left(\frac{s}{s_0}\right)^{\!\!0.079}\,\mb\ .
\label{145}
 \EE

The dipole cross section Eqs.(\ref{130}) -- (\ref{145}) provides the
imaginary part of the elastic amplitude.  It is known, however, that the
energy dependence of the total cross section generates also a real part
\cite{bronzan},
 \beq
\sigma_{\bar qq}(r,s) \Rightarrow
\left(1-i\,\frac{\pi}{2}\,\frac{\partial}
{\partial\ln(s)}\right)\,
\sigma_{\bar qq}(r,s)
\label{real-part}
 \eeq
 The energy dependence of the dipole cross section Eq.~(\ref{130}) is
rather steep at small $r$ leading to a large real part which should not
be neglected. For instance, the photoproduction amplitude of $\gamma N\to
J/\Psi N$ rises $\propto s^{0.2}$ and the real-to-imaginary part ratio is
over $30\%$. 

Although the
form of Eq.~(\ref{130}) successfully describes data for DIS at small $x$
only up to $Q^2\approx 10\GeV^2$ we prefer this parameterization for study 
of charmonium electroproduction.
The reason is 
that we want to study CT effects predominantly in the range $Q^2\leq
20\GeV^2$ and in addition 
parameterization Eq.~(\ref{130}) 
describes the transition toward photoproduction limit 
better than parameterization presented in \cite{gbw}.
Besides, in the paper \cite{hikt-00}
was shown studying electroproduction of charmonia
off nucleons  that the difference between
predictions using both parameterizations \cite{gbw} and Eq.~(\ref{130})
is rather small and can be taken
as a measure of the theoretical uncertainty.
\vspace*{1.0cm}

The perturbative distribution amplitude (``wave function'') of the $\bar
qq$ ($\bar cc$ for $J/\Psi$ production) Fock component of the photon 
has the following form 
for transversely (T) and longitudinally (L) polarized photons 
\cite{lc,bks-71,nz-91},
 \BE
\Psi_{\bar qq}^{T,L}({\vec{r}},\alpha) =
\frac{\sqrt{N_{C}\,\alpha_{em}}}{2\,\pi}\,\,
Z_{q}\,\bar{\chi}\,\hat{O}^{T,L}\,\chi\, 
K_{0}(\epsilon\,r)
\label{70}
 \EE
 where $\chi$ and $\bar{\chi}$ are the spinors of the quark and
antiquark, respectively; $Z_{q}$ is the quark charge,
$Z_{q} = Z_{c} = 2/3$ for $J/\Psi$ production; $N_{C} = 3$ is the
number of colors. $K_{0}(\epsilon r)$ is a modified Bessel
function with 
 \BE
\epsilon^{2} =
\alpha\,(1-\alpha)\,Q^{2} + m_{c}^{2}\ ,
\label{80}
 \EE
 where $m_{c} = 1.5\GeV$ is mass of the $c$ quark, and $\alpha$ is the 
fraction of the LC momentum of the photon carried by the quark. The operators
$\widehat{O}^{T,L}$ read,
 \BE 
\widehat{O}^{T} = m_{c}\,\,\vec{\sigma}\cdot\vec{e} +
i\,(1-2\alpha)\,(\vec{\sigma}\cdot\vec{n})\,
(\vec{e}\cdot\vec{\nabla}_r) + (\vec{\sigma}\times
\vec{e})\cdot\vec{\nabla}_r\ ,
 \label{90}
 \EE
 \BE
\widehat{O}^{L} =
2\,Q\,\alpha (1 - \alpha)\,(\vec{\sigma}\cdot\vec{n})\ .
\label{100}
 \EE
 Here $\vec\nabla_r$ acts on transverse coordinate $\vec r$;
$\vec{e}$ is the polarization vector of the photon and $\vec{n}$ is a unit
vector parallel to the photon momentum.

In general,
the transverse $\bar qq$ separation is controlled by the distribution 
amplitude Eq.~(\ref{70}) with the mean value,
 \BE
\la r\ra \sim \frac{1}{\epsilon} = 
\frac{1}{\sqrt{Q^{2}\,\alpha\,(1-\alpha) + m_{q}^{2}}}\,.
\label{212}
 \EE

For production of light vector meson
very asymmetric $\bar qq$ pairs with $\alpha$ or $(1-\alpha) \lsim
m_q^2/Q^2$ become possible. Consequently,
the mean transverse separation $\la r\ra \sim 1/m_q$ becomes
huge since one must use current quark masses within pQCD. 
However, that is not the case in charmonium production because of a large
quark mass $m_c = 1.5\GeV$.
Therefore, we are out of the problem how to include
nonperturbative interaction effects between $c$ and $\bar c$
because they are rather small.
Despite of this fact for completeness
we include these nonperturbative interaction effects 
in all calculations to avoid  
although small but supplementary
uncertainties in predictions. 
We take from \cite{kst2} the corresponding phenomenology including the
interaction between $c$ and $\bar c$  based on the light-cone Green function
approach.

The Green function $G_{\bar qq}(z_1,\vec r_1;z_2,\vec r_2)$ 
describes the 
propagation of an interacting $\bar qq$ pair 
($\bar cc$ pair for the case of $J/\Psi$ production)
between points with
longitudinal coordinates $z_{1}$ and $z_{2}$ and with initial and final
separations $\vec r_1$ and $\vec r_2$. This
Green function satisfies the 
two-dimensional Schr\"odinger equation, 
 \BE
i\frac{d}{dz_2}\,G_{\bar qq}(z_1,\vec r_1;z_2,\vec r_2)=
\left[\frac{\epsilon^{2} - \Delta_{r_{2}}}{2\,\nu\,\alpha\,(1-\alpha)}
+V_{\bar qq}(z_2,\vec r_2,\alpha)\right]
G_{\bar qq}(z_1,\vec r_1;z_2,\vec r_2)\ .
\label{250}  
 \EE
 Here $\nu$ is the photon energy. The Laplacian $\Delta_{r}$ acts on
the coordinate $r$.  

The imaginary part of the LC potential $V_{\bar
qq}(z_2,\vec r_2,\alpha)$ in (\ref{250}) is responsible for
attenuation of the $\bar qq$ in the medium, while the real part
represents the interaction between the $q$ and $\bar{q}$.  
This potential is supposed to provide the correct LC wave functions of 
vector mesons. For the sake of simplicity we use  the oscillator form 
of the potential,
 \BE  
{\rm Re}\,V_{\bar qq}(z_2,\vec r_{2},\alpha) =
\frac{a^4(\alpha)\,\vec r_{2}\,^2} 
{2\,\nu\,\alpha(1-\alpha)}\ ,
\label{260} 
 \EE
 which leads to a Gaussian $r$-dependence of the LC wave function of the
meson ground state.  The shape of the function $a(\alpha)$ will be
discussed below.

 In this case equation (\ref{250}) has an analytical solution, the
harmonic oscillator Green function \cite{fg},
 \BA 
G_{\bar qq}(z_1,\vec r_1;z_2,\vec r_2) =
\frac{a^2(\alpha)}{2\;\pi\;i\;
{\rm sin}(\omega\,\Delta z)}\, {\rm exp}
\left\{\frac{i\,a^2(\alpha)}{{\rm sin}(\omega\,\Delta z)}\,
\Bigl[(r_1^2+r_2^2)\,{\rm cos}(\omega \;\Delta z) -
2\;\vec r_1\cdot\vec r_2\Bigr]\right\}
\nonumber\\ \times \,
{\rm exp}\left[- 
\frac{i\,\epsilon^{2}\,\Delta z}
{2\,\nu\,\alpha\,(1-\alpha)}\right] \ , 
\label{270} 
 \EA
where $\Delta z=z_2-z_1$ and 
 \BE \omega = \frac{a^2(\alpha)}{\nu\;\alpha(1-\alpha)}\ .
\label{280} 
 \EE
 The boundary condition is $G_{\bar
qq}(z_1,\vec r_1;z_2,\vec r_2)|_{z_2=z_1}=
\delta^2(\vec r_1-\vec r_2)$.

The probability amplitude to find the $\bar qq$ fluctuation of a photon
at the point $z_2$ with separation $\vec r$ is given by an integral
over the point $z_1$ where the $\bar qq$ is created by the photon with
initial separation zero,
 \BE
\Psi^{T,L}_{\bar qq}(\vec r,\alpha)=
\frac{i\,Z_q\sqrt{\alpha_{em}}}
{4\pi\,\nu\,\alpha(1-\alpha)} 
\int\limits_{-\infty}^{z_2}dz_1\,
\Bigl(\bar\chi\;\widehat O^{T,L}\chi\Bigr)\,
G_{\bar qq}(z_1,\vec r_1;z_2,\vec r)
\Bigr|_{r_1=0}\ .
\label{290}
 \EE
 The operators $\widehat O^{T,L}$ are defined in Eqs.~(\ref{90}) and
(\ref{100}). Here they act on the coordinate $\vec r_1$.

If we write the transverse part  as
 \BE
\bar\chi\;\widehat O^{T}\chi= A+\vec B\cdot\vec\nabla_{r_1}\ ,
\label{300}
 \EE
 then the distribution functions read,   
 \BE
\Psi^{T}_{\bar qq}(\vec r,\alpha) =
Z_q\sqrt{\alpha_{em}}\,\left[A\,\Phi_0(\epsilon,r,\lambda)
+ \vec B\,\vec\Phi_1(\epsilon,r,\lambda)\right]\ ,
\label{310}
 \EE
 \BE
\Psi^{L}_{\bar qq}(\vec r,\alpha) =
2\,Z_q\sqrt{\alpha_{em}}\,Q\,\alpha(1-\alpha)\,
\bar\chi\;\vec\sigma\cdot\vec n\;\chi\,
\Phi_0(\epsilon,r,\lambda)\ ,
\label{320}
 \EE
 where
 \BE
\lambda=
\frac{2\,a^2(\alpha)}{\epsilon^2}\ .
\label{330}
 \EE

The functions $\Phi_{0,1}$ in Eqs.~(\ref{310}) and (\ref{320})
are defined as
 \BE
\Phi_0(\epsilon,r,\lambda) =
\frac{1}{4\pi}\int\limits_{0}^{\infty}dt\,
\frac{\lambda}{{\rm sh}(\lambda t)}\,
{\rm exp}\left[-\ \frac{\lambda\epsilon^2 r^2}{4}\,
{\rm cth}(\lambda t) - t\right]\ ,
\label{340}
 \EE
 \BE
\vec\Phi_1(\epsilon,r,\lambda) =
\frac{\epsilon^2\vec r}{8\pi}\int\limits_{0}^{\infty}dt\,
\left[\frac{\lambda}{{\rm sh}(\lambda t)}\right]^2\,
{\rm exp}\left[-\ \frac{\lambda\epsilon^2 r^2}{4}\,
{\rm cth}(\lambda t) - t\right]\ .  
\label{350}
 \EE

Note that the $\bar q-q$ interaction enters Eqs.~(\ref{310}) and
(\ref{320}) via the parameter $\lambda$ defined in (\ref{330}). In the
limit of vanishing interaction $\lambda\to 0$ (i.e. $Q^2\to \infty$,
$\alpha$ is fixed, $\alpha\not=0$ or $1$) Eqs.~(\ref{310}) -
(\ref{320})  produce the perturbative expressions of Eq.~(\ref{70}).
As was mentioned above, for charmonium production 
nonperturbative interaction effects are quite weak. Consequently, 
the parameter $\lambda$ is rather small due to a large quark mass 
$m_c = 1.5\GeV$ :
 \BE
\lambda=
\frac{8\,a^2(\alpha)}{Q^2 + 4~m_{c}^{2}}\ .
\label{355}
 \EE

With the choice $a^2(\alpha)\propto \alpha(1-\alpha)$ the end-point
behavior of the mean square interquark separation $\la r^2\ra\propto
1/\alpha(1-\alpha)$ contradicts the idea of confinement. Following
\cite{kst2} we fix this problem via a simple modification of the LC
potential,
 \BE
a^2(\alpha) = a^2_0 +4a_1^2\,\alpha(1-\alpha)\ .
\label{180}
 \EE 
 The parameters $a_0$ and $a_1$ were adjusted in \cite{kst2} to data on
total photoabsorption cross section \cite{gamma1,gamma2}, diffractive
photon dissociation and shadowing in nuclear photoabsorption reaction.  
The results of our calculations vary within only $1\%$ when $a_0$ and
$a_1$ satisfy the relation,
 \BA
a_0^2&=&v^{ 1.15}\, (0.112)^2\,\GeV^{2}\nonumber\\
a_1^2&=&(1-v)^{1.15}\,(0.165)^2\,\GeV^{2}\ , 
\label{190} 
 \EA
 where $v$ takes any value $0<v<1$. In view of this insensitivity of the
observables we fix the parameters at $v=1/2$. 
We checked that this choice does not affect our results beyond a
few percent uncertainty.
\vspace*{1.0cm}

The last ingredient in elastic production amplitude 
(\ref{120}) is charmonium wave function.
We use a popular prescription
\cite{terentev} applying the Lorentz boost to the rest frame wave
function assumed to be Gaussian which leads to radial parts of
transversely and longitudinally polarized mesons in the form,
 \BE
\Phi_{J/\Psi}^{T,L}(\vec r,\alpha) = 
C^{T,L}\,\alpha(1-\alpha)\,f(\alpha)\,
{\rm exp}\left[-\ \frac{\alpha(1-\alpha)\,{\vec r}^2}
{2\,R^2}\right]\  
\label{170}
 \EE
 with a normalization defined below, and 
 \beq
f(\alpha) = \exp\left[-\ \frac{m_c^2\,R^2}
{2\,\alpha(1-\alpha)}\right]\ 
\label{170a}
 \eeq 
with the parameters from \cite{jan97}, $R=0.183\,\fm$ and
$m_c=1.3\GeV$.
A detailed analysis of various problems in
this relativization procedure \cite{hz} leads to the same form (\ref{170}).

We assume that the distribution amplitude of $\bar cc$ fluctuations for
$J/\Psi$ and for the photon have a similar structure
\cite{jan97}.  Then in analogy to Eqs.~(\ref{310}) -- (\ref{320}),
 \beqn
\Psi^T_{J/\Psi}(\vec r,\alpha) &=&
(A+\vec B\cdot\vec\nabla)\,
\Phi^T_{J/\Psi}(r,\alpha)\ ;
\label{172}\\
\Psi^L_{J/\Psi}(\vec r,\alpha) &=& 
2\,m_{J/\Psi}\,\alpha(1-\alpha)\,
(\bar\chi\,\vec\sigma\cdot\vec n\,\chi)\,
\Phi^L_{J/\Psi}(r,\alpha)\ .
\label{174}
 \eeqn

Correspondingly, the normalization conditions for the transverse 
and longitudinal charmonium wave
functions read,
 \BE
N_{C}\,\int d^{2} r\,\int d\alpha \left\{ m_{c}^{2}\,
\Bigl|\Phi^T_{J/\Psi}(\vec r,\alpha)\Bigr|^{2} + \Bigl[\alpha^{2} + 
(1-\alpha)^{2}\Bigr]\,
\Bigl|\partial_{r}
\Phi^T_{J/\Psi}(\vec r,\alpha)\Bigr|^{2} \right\} = 1
\label{230}
 \EE
 \BE
4\,N_{C}\,\int d^{2} r\,\int d\alpha\, 
\alpha^{2}\,(1-\alpha)^{2}\,
m_{J/\Psi}^{2}\,\Bigl|\Phi^L_{J/\Psi}(\vec r,\alpha)\Bigr|^2 = 1\, .
\label{240}
 \EE

\section{Electroproduction of $J/\Psi$ on a nucleon, comparison 
with data}
\label{data-N}

In this section we verify first 
the LC approach by comparing with data for nucleon target. 
The forward production amplitude
$\gamma^*\,N \to J/\Psi\,N$ for transverse and longitudinal photons and 
charmonium is calculated
using the nonperturbative photon Eqs.~(\ref{310}),
(\ref{320}) and vector meson Eqs.~(\ref{172}), (\ref{174}) wave functions.  
and has the following form,
 \BA
{\cal M}_{\gamma^{*}N\rightarrow J/\Psi\,N}^{T}(s,Q^{2})
\Bigr|_{t=0} &=& 
N_{C}\,Z_{c}\,\sqrt{\alpha_{em}}
\int d^{2} r\,\sigma_{\bar qq}(\vec r,s)
\int\limits_0^1 d\alpha \Bigl\{ m_{c}^{2}\,
\Phi_{0}(\epsilon,\vec r,\lambda)\Psi^T_{J/\Psi}(\vec r,\alpha)
\nonumber\\ 
&+& \bigl [\alpha^{2} + (1-\alpha)^{2}\bigr ]\,
\vec{\Phi}_{1}(\epsilon,\vec r,\lambda)\cdot
\vec{\nabla}_{r}\,\Psi^T_{J/\Psi}(\vec r,\alpha) \Bigr\}\,;
\label{360}
 \EA
 \BA
{\cal M}_{\gamma^{*}N\rightarrow J/\Psi\,N}^{L}(s,Q^{2})
\Bigr|_{t=0} &=& 
4\,N_{C}\,Z_{c}\,\sqrt{\alpha_{em}}\,m_{J/\Psi}\,Q\,
\int d^{2} r\,\sigma_{\bar qq}(\vec r,s)
\nonumber\\&\times& 
\int\limits_0^1 d\alpha\,
\alpha^{2}\,(1-\alpha)^{2}\, 
\Phi_{0}(\epsilon,\vec r,\lambda)
\Psi^L_{J/\Psi}(\vec r,\alpha)\ .
\label{370}
 \EA
 These amplitudes are normalized as ${|{\cal M}^{T,L}|^{2}}=
\left.16\pi\,{d\sigma_{N}^{T,L}/ dt}\right|_{t=0}$. The real
part of the amplitude is included 
according to the prescription described in the previous section.
We calculate the cross sections
$\sigma = \sigma^T + \epsilon\,\sigma^L$ assuming that the photon
polarization is $\epsilon=1$.
 \begin{figure}[bht]
\includegraphics{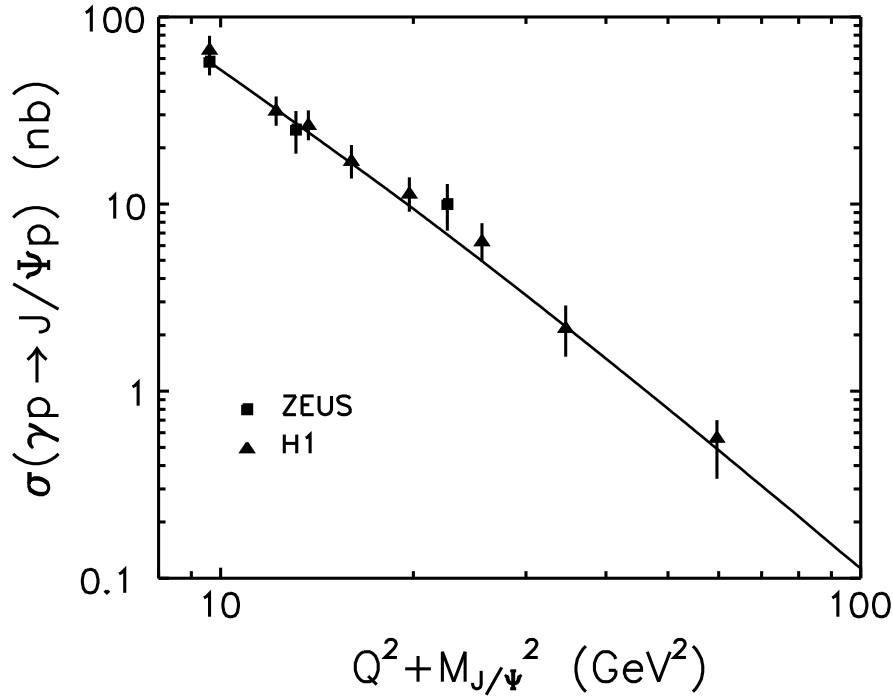}
\begin{center}
\vspace{9.0cm}
\parbox{13cm}
{\caption[Delta]
 {\sl $Q^2+m_{J/\Psi}^{2}$- dependence of the integrated cross section for the 
reactions $\gamma^*\,p \to J/\Psi\,p$.  
The model calculations are compared with H1
\cite{H1-jpsi-q2} and ZEUS 
\cite{ZEUS-jpsi-q2} data at energy $W=90\GeV$.
}
 \label{q2-nucl}}
\end{center}
 \end{figure}

Now we can check the absolute value of the production cross section by
comparing with data for elastic charmonium electroproduction 
$\gamma^*\,p \to J/\Psi\,p$.  
Unfortunately, data are available only for
the cross section integrated over $t$, 
 \beq
\sigma^{T,L}(\gamma^{*}N\to VN) = 
\frac{|{\cal M}^{T,L}|^{2}}
{16\pi\,B_{\gamma^*N}}\ ,
\label{375}
 \eeq
 where $B$ is the slope parameter in reaction $\gamma^*\,p \to J/\Psi\,p$.
We use the experimental value \cite{H1-jpsi-w} $B=4.7\GeV^{-2}$. 

Our predictions are plotted in Fig.~\ref{q2-nucl} together with the data
on the $Q^2+m_{J/\Psi}^{2}$ dependence of the cross section from H1 
\cite{H1-jpsi-q2} and ZEUS \cite{ZEUS-jpsi-q2}.

 \begin{figure}[thb]
\includegraphics{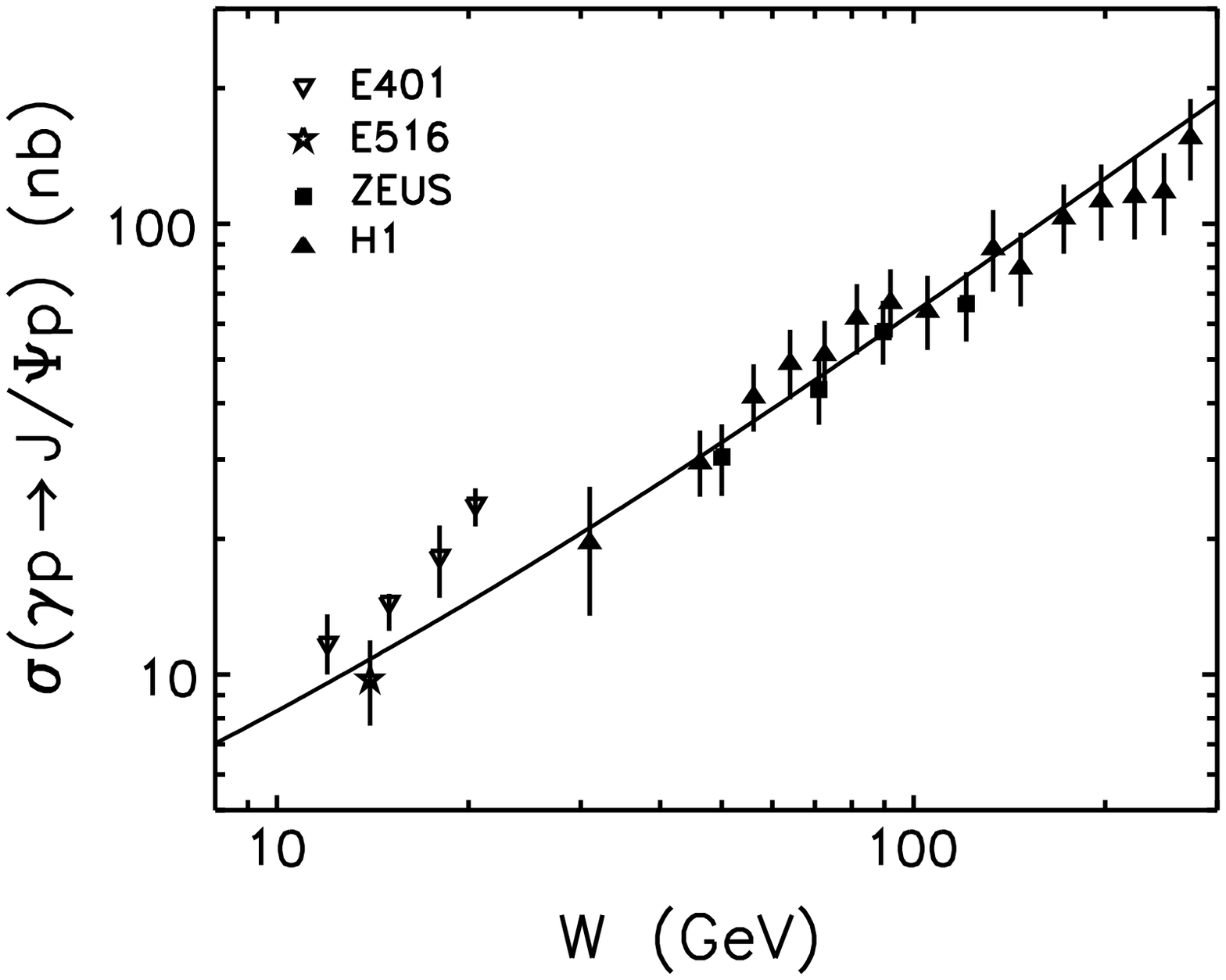}
\begin{center}
\vspace{9.0cm}
\parbox{13cm}
{\caption[Delta]
{\sl Energy dependence of the real photoproduction cross section
on a nucleon, $\gamma\,p \to J/\Psi\,p$. Our results are 
compared with data from the fixed target E401
\cite{E401-jpsi-w}, E516 \cite{E516-jpsi-w}, and collider HERA H1 
\cite{H1-jpsi-w} and 
ZEUS \cite{ZEUS-jpsi-w} experiments.}
\label{real-nucl}}
\end{center}
 \end{figure}

As the second test of
our approach is a
description of the low $Q^2$ region as well.
As we discussed in the previous section despite 
of smallness of effects of nonperturbative interaction 
between the $c$ and $\bar c$,
we include them into calculations.
Comparison of the model
with data \cite{E401-jpsi-w,E516-jpsi-w,H1-jpsi-w,ZEUS-jpsi-w}
for the energy dependence of the cross section of real $J/\Psi$
photoproduction is presented in Fig.~\ref{real-nucl}.

The normalization of the cross section and its energy and $Q^2$
dependence are remarkably well reproduced in Figs.~\ref{q2-nucl} --
\ref{real-nucl}. This is an important achievement since the absolute
normalization is usually much more difficult to reproduce 
the production cross sections than nuclear
effects. For instance, the similar, but simplified calculations in
\cite{kz-91} underestimate the $J/\Psi$ photoproduction cross section on
protons by an order of magnitude.

As a cross-check for the choice of the $J/\Psi$ wave function in 
Eqs.~(\ref{170}) and (\ref{170a}) we also calculated the total
$J/\Psi$-nucleon cross section, which was already estimated in
\cite{hikt-00} using the charmonium wave functions calculated
with several realistic $\bar qq$ potentials.
The $J/\Psi$-nucleon total cross section has the form,
 \BA
\sigma_{tot}^{J/\Psi N} = 
N_{C}\,\int d^{2} r\,
\int d\alpha \left\{ m_{c}^{2}\,
\Bigl|\Phi^T_{J/\Psi} (\vec r,\alpha)\Bigr |^2 + 
\bigl [\alpha^{2} + (1-\alpha)^{2}\bigr ]\,
\Bigl |\partial_{r}\Phi^T_{J/\Psi}(\vec r,\alpha)\Bigr|^{2} 
\right\}\,
\sigma_{\bar qq}(\vec r,s)
\label{380}
 \EA
 We calculated $\sigma_{tot}^{J/\Psi N}$ with the charmonium wave
function in the form (\ref{170}) with the parameters described in the
previous section. For the dipole cross section we adopt the KST
parameterization (\ref{130}) which is designed to describe low-$Q^2$
data. Then, at $\sqrt{s} = 10\GeV$ we obtain $\sigma_{tot}^{J/\Psi N} =
4.2\,\mb$ which is in a good agreement with 
$\sigma_{tot}^{J/\Psi N} = 3.6\,\mb$ evaluated in \cite{hikt-00}.

\section{Incoherent production of charmonia off nuclei}
\label{psi-incoh}

\subsection{the LC Green function formalism}

In this section we present a short review of the LC Green function
formalism for incoherent production of an arbitrary vector meson.
For the case of charmonium production one should replace $V\rightarrow
J/\Psi$ and $\bar qq\rightarrow\bar cc$.
In general,
the diffractive incoherent (quasielastic) production of vector mesons
off nuclei, $\gamma^{*}\,A\rightarrow V\,X$, 
is associated with a break up of the nucleus, but without production
of new particles.
In another words,
one sums over all final states of the target nucleus except those which 
contain particle (pion) creation. 
The observable usually studied experimentally is nuclear transparency 
defined as 
 \BE
Tr^{inc}_{A} = 
\frac{\sigma_{\gamma^{*}A\to VX}^{inc}}
{A\,\sigma_{\gamma^{*}N\to VN}}\ .
\label{480}
 \EE
 The $t$-slope of the differential quasielastic cross section is the same
as on a nucleon target. Therefore, instead of integrated cross sections
one can also use nuclear transparency expressed via
the forward differential cross sections 
Eq.~(\ref{125}),
 \beq 
Tr^{inc}_A = \frac{1}{A}\,
\left|\frac{{\cal M}_{\gamma^{*}A\to VX}(s,Q^{2})}
{{\cal M}_{\gamma^{*}N\to VN}(s,Q^{2})}\right|^2\, .
\label{485}
 \eeq 

In the LC Green function approach \cite{knst-01}
the physical photon 
$|\gamma^*\ra$ is decomposed 
into different Fock states, namely, the bare photon
$|\gamma^*\ra_0$, $|\bar qq\ra$, $|\bar qqG\ra$, etc. 
As we mentioned above the higher Fock states
containing gluons describe the energy dependence of the
photoproduction reaction on a nucleon. 
Besides, those Fock components also lead to gluon shadowing  
as far as nuclear effects are concerned.
However, these fluctuations are heavier and have a
shorter coherence time (lifetime) than the lowest $|\bar qq\ra$
state. Therefore, at medium energies only $|\bar qq\ra$ fluctuations of 
the photon matter. Consequently, gluon shadowing related to the higher Fock
states will be dominated at high energies.
Detailed description and calculation of gluon shadowing 
for the case of vector meson production off nuclei is presented
in \cite{knst-01,ikth-02}.
Although the gluon shadowing is quite small in the kinematic range
for study of CT effects suggested 
in the present paper we include it in all the calculations.

Propagation of an interacting $\bar qq$ pair in a nuclear medium is also
described by the Green function satisfying the evolution Eq.~(\ref{250}).
However, the potential in this case acquires an imaginary part which
represents absorption in the medium (see (\ref{40}) for notations),
 \BE
Im V_{\bar qq}(z_2,\vec r,\alpha) = - 
\frac{\sigma_{\bar qq}(\vec r,s)}{2}\,\rho_{A}({b},z_2)\,.
\label{440}
 \EE
 The evolution equation (\ref{250}) with the potential
$V_{\bar qq}(z_{2},\vec r_{2},\alpha)$ containing this imaginary 
part
was used in \cite{krt1,krt2}, and nuclear shadowing in deep-inelastic
scattering was calculated in good agreement with data.

The analytical solution of Eq.~(\ref{270}) is only known for the 
harmonic oscillator potential $V(r)\propto r^2$. To keep
the calculations reasonably simple we are forced to use the dipole 
approximation
 \beq
\sigma_{\bar qq}(r,s) = C(s)\,r^2\ ,
\label{460}
 \eeq
which allows to obtain the Green function in an analytical form.

The energy dependent factor $C(s)$ is adjusted 
by demanding that calculations employing the
approximation Eq.~(\ref{460}) reproduce correctly the results based on
the realistic cross section in the limit $l_c\gg R_A$ 
(the so called ``frozen'' approximation)
when the Green function takes the simple form : 
 \beq
G_{\bar qq}(z_1,\vec r_1;z_2,\vec r_2) \Rightarrow
\delta(\vec r_1-\vec r_2)\,\exp\left[
-{1\over2}\,\sigma_{\bar qq}(r_1)
\int\limits_{z_1}^{z_2} dz\,\rho_A(b,z)\right]\ ,
\label{465}
 \eeq 
 where the dependence of the Green function on impact parameter is
dropped. 
Thus, for incoherent
production of vector mesons the factor $C(s)$ is fixed by the relation,
 \beqn
&& \frac{
\int d^{2}{b}\,T_A(b)\,\left|
\int d^{2}r\,r^2\,\exp\biggl[ - {1\over2}\, 
C_{T,L}(s)\,r^2\,T_{A}({b})\,\biggr]
\int d\alpha\,
\Psi_{V}^{*}(\vec r,\alpha)\,
\Psi^{T,L}_{\gamma^{*}}(\vec r,\alpha)
\right|^2 }
{\left|\int d^{2}r\,r^2\int d\alpha\,
\Psi_{V}^{*}(\vec r,\alpha)\,
\Psi^{T,L}_{\gamma^{*}}(\vec r,\alpha)\right|^2 }  
\nonumber\\ &=&
\frac{
\int d^{2}{b}\,T_A(b)\,\left|
\int d^{2}r\,\sigma_{\bar qq}(r,s)\,
\exp\biggl[ - {1\over2}\, 
\sigma_{\bar qq}(r,s)\,T_{A}({b})\,\biggr]
\int d\alpha\,
\Psi_{V}^{*}(\vec r,\alpha)\,
\Psi^{T,L}_{\gamma^{*}}(\vec r,\alpha)
\right|^2 }
{\left|\int d^{2}r\,\sigma_{\bar qq}(r,s)\int d\alpha\,
\Psi_{V}^{*}(\vec r,\alpha)\,
\Psi^{T,L}_{\gamma^{*}}(\vec r,\alpha)\right|^2 } 
\label{467}
 \eeqn

To take advantage of the analytical form of the Green function which is
known only for the LC potential Eq.~(\ref{440}) with a constant nuclear
density, we use the approximation $\rho_{A}({b},z)  =
\rho_{0}\,\Theta(R_{A}^{2} - {b}^{2} - z^{2})$. Therefore we have to use
this form for Eq.~(\ref{467}) as well. The value of the
mean nuclear density $\rho_{0}$ has been determined using the relation,
 \BE
\int d^{2}{b}\,\biggl [ 1 - exp \biggl ( - \sigma_{0}\,
\rho_{0}\,\sqrt{R_{A}^{2} - {b}^{2}}\biggr ) \biggr ] =
\int d^{2}{b}\,\biggl [ 1 - exp 
\biggl ( - \frac{\sigma_{0}}{2}\,T({b})
\biggr ) \biggr ]\ ,
\label{469}
 \EE
 where the nuclear thickness function $T_A(b)$ is calculated with the
realistic Wood-Saxon form of the nuclear density. The value of $\rho_{0}$
turns out to be practically independent of the cross section $\sigma_{0}$
in the range from 1 to 50 mb.

With the potential Eqs.~(\ref{440}) -- (\ref{460}) the solution of 
Eq.~(\ref{250}) has the same form as Eq.~(\ref{270}), except that one 
should 
replace $\omega \Rightarrow \Omega$, where
 \beq
\Omega = \frac{\sqrt{a^4(\alpha)-
i\,\rho_{A}({b},z)\,
\nu\,\alpha\,(1-\alpha)\,C(s)}}
{\nu\;\alpha(1-\alpha)}\ .
\label{470}
 \eeq

As we discussed in \cite{knst-01} the value of $l_c$ can 
distinguish different regimes of vector meson production.

{\bf (i)} The CL is much shorter than the mean nucleon spacing in a
nucleus ($l_c \to 0$). In this case $G(z_2,\vec r_2;z_1,\vec r_1)  \to
\delta(z_2-z_1)$. Correspondingly, the formation time of the
meson wave function is very short as well as given in
Eq.~(\ref{20}). For light vector mesons $l_f\sim l_c$ and 
since formation and coherence lengths are proportional to
photon energy both must be short. 
Consequently, nuclear transparency is given by the
simple formula Eq.~(\ref{40}) corresponding to the Glauber
approximation.

{\bf (ii)}
In production of charmonia and other heavy flavor quarkonii
the intermediate case $l_c\to 0$, but $l_f\sim R_A$ can  
be realized. Then
the formation of the meson wave function 
is described by the Green function and the numerator of the nuclear 
transparency ratio Eq.~(\ref{485}) has the form 
\cite{kz-91},
 \beq
\Bigl|{\cal M}_{\gamma^{*}A\to VX}(s,Q^{2})
\Bigr|^2_{l_c\to0;\,l_f\sim R_A} = 
\int d^2b\int_{-\infty}^{\infty} dz\,\rho_A(b,z)\,
\Bigl|F_1(b,z)\Bigr|^2\ ,
\label{500}
 \eeq
 where
 \beq
F_1(b,z) = 
\int_0^1 d\alpha
\int d^{2} r_{1}\,d^{2} r_{2}\,
\Psi^{*}_{V}(\vec r_{2},\alpha)\,
G(z^\prime,\vec r_{2};z,\vec r_{1})\,
\sigma_{\bar qq}(r_{1},s)\,
\Psi_{\gamma^{*}}(\vec r_{1},
\alpha)\Bigl|_{z^\prime\to\infty}
\label{505}
 \eeq

{\bf (iii)} 
In the high energy limit
$l_c \gg R_A$ (in fact, it is more correct to compare with
the mean free path of the $\bar qq$ in a nuclear medium if the latter is 
shorter
than the nuclear radius). 
In this case $G(z_2,\vec r_2;z_1,\vec r_1)
\to \delta(\vec r_2 - \vec r_1)$, i.e. all fluctuations of 
the transverse $\bar qq$ 
separation are ``frozen'' by Lorentz time dilation.
Then, the numerator on the r.h.s. of Eq.~(\ref{485}) takes the form
\cite{kz-91},
 \beqn
\Bigl|{\cal M}_{\gamma^{*}A\to VX}(s,Q^{2})
\Bigr|^2_{l_c \gg R_A} &=& 
\int d^2b\,T_A(b)\left|\int d^2r\int_0^1 d\alpha 
\right. \label{510}\\
&\times& \left.\Psi_{V}^{*}(\vec r,\alpha)\,
\sigma_{\bar qq}(r,s)\,  
\exp\left[-{1\over2}\sigma_{\bar qq}(r,s)\,T_A(b)\right]
\Psi_{\gamma^{*}}(\vec r,\alpha,Q^2)\right|^2\ .
\nonumber
 \eeqn 
 In this case the $\bar qq$ attenuates with a constant absorption cross
section like in the Glauber model, except that the whole exponential is
averaged rather than just the cross section in the exponent. The
difference between the results of the two prescriptions are the well 
known inelastic corrections of Gribov \cite{zkl}.

{\bf (iv)} 
This regime reflects the general case when there is
no restrictions for either $l_c$ or $l_f$.  
The corresponding theoretical tool 
has been developed for the first time only recently in \cite{knst-01}
and applied to electroproduction of light vector mesons
at medium and high energies.
Even within the VDM the Glauber model expression interpolating between
the limiting cases of low [(i), (ii)] and high [(iii)] energies has been
derived only recently \cite{hkn} as well.
In this general case
the incoherent photoproduction amplitude is
represented as a sum of two terms \cite{hkz},
 \BE 
\Bigl|\,{\cal M}_{\gamma^{*}A\to
VX}(s,Q^{2})\Bigr|^{2} = \int d^{2}b
\int\limits_{-\infty}^{\infty} dz\,\rho_{A}({b},z)\, 
\Bigl|F_{1}({b},z) - F_{2}({b},z)\Bigr|^{2}\ .
\label{520}
 \EE
 The first term $F_{1}({b},z)$ introduced above in Eq.~(\ref{505}) 
alone would correspond to the short
$l_c$ limit (ii). The second term $F_{2}({b},z)$ in (\ref{520})  
corresponds to the situation when the
incident photon produces a $\bar qq$ pair diffractively and coherently at
the point $z_1$ prior to incoherent quasielastic scattering at point $z$.
The LC Green functions describe the evolution of the $\bar qq$ over the
distance from $z_1$ to $z$ and further on, up to the formation of the
meson wave function. Correspondingly, this term has the form,
 \beqn
F_{2}(b,z) &=& \frac{1}{2}\,
\int\limits_{-\infty}^{z} dz_{1}\,\rho_{A}(b,z_1)\,
\int\limits_0^1 d\alpha\int d^2 r_1\,
d^2 r_{2}\,d^2 r\,
\Psi^*_V (\vec r_2,\alpha)
\nonumber \\
&\times&
G(z^{\prime}\to\infty,\vec r_2;z,\vec r)\,
\sigma_{\bar qq}(\vec r,s)\,
G(z,\vec r;z_1,\vec r_1)\,
\sigma_{\bar qq}(\vec r_1,s)\,
\Psi_{\gamma^{*}}(\vec r_1,\alpha)\, .
\label{530}
 \eeqn

Eq.~(\ref{520}) correctly reproduces the limits (i) - (iii). 
At $l_c\to 0$ the second term $F_2(b,z)$ vanishes because of strong
oscillations, and Eq.~(\ref{520}) reproduces the Glauber expression
Eq.~(\ref{40}). 
At $l_c\gg R_A$ the phase shift in the
Green functions can be neglected and they acquire the simple form
$G(z_2,\vec r_2;z_1,\vec r_1) \to \delta(\vec r_2 - \vec r_1)$. In this
case the integration over longitudinal coordinates in Eqs.~(\ref{505})
and (\ref{530}) can be performed explicitly and the asymptotic expression
Eq.~(\ref{510}) is recovered as well.

\subsection{Comparison with data for incoherent production 
of $J/\Psi$}\label{incoh-data}

Exclusive incoherent electroproduction of vector mesons off nuclei has
been suggested in \cite{knnz} to be very convenient for investigation of CT. 
Increasing
the photon virtuality $Q^2$ one squeezes the produced $\bar qq$ wave
packet. Such a small colorless system propagates through the nucleus with
little attenuation, provided that the energy is sufficiently high
($l_f\gg R_A$) so the fluctuations of the $\bar qq$ separation are frozen 
during propagation.
Consequently, a rise of nuclear transparency
$Tr_A^{inc}(Q^2)$ with $Q^2$ should give a signal for CT.
Indeed, such a rise was observed in the E665 experiment \cite{e665-rho}
at Fermilab for exclusive production of $\rho^0$ mesons off
nuclei what has been claimed as manifestation of CT.

However, the effect of coherence length \cite{kn95,hkn} leads also
to a rise of $Tr_A^{inc}(Q^2)$ with $Q^2$ and so can imitate
CT effects.
This happens when the coherence length varies from long to short 
(see Eq.~(\ref{30}))
compared to the nuclear size and  
the length of the path in nuclear matter becomes shorter.
Consequently, the vector meson (or $\bar qq$)
attenuates less in nuclear medium.
This happens when $Q^2$ increases at fixed $\nu$.
Therefore one should carefully disentangle these two phenomena.
 \begin{figure}[htb]
\includegraphics{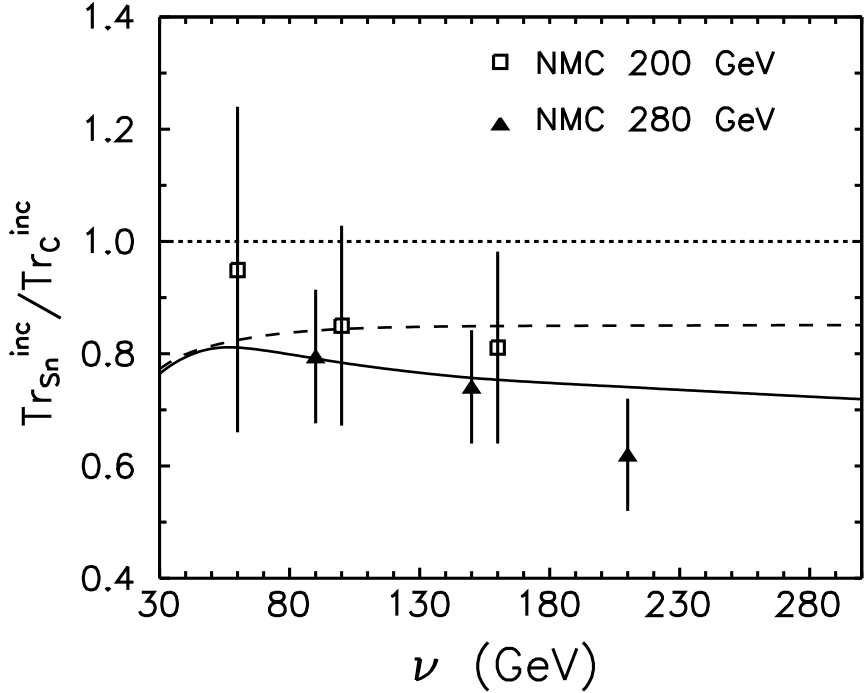}
\begin{center}
\vspace{9.0cm}
\parbox{13cm}
{\caption[Delta]
{ 
Energy dependence of the ratio of nuclear transparencies 
$Tr_{Sn}$ and $Tr_{C}$ vs. 
experimental points taken from the NMC experiment 
\cite{NMC-psi}. 
Solid and dashed curves show our results using the LC Green function 
approach in general case with no restriction 
for either $l_c$ or $l_f$ Eq.~(\ref{520}) and in the limit of 
$l_c\to 0$  Eq.~(\ref{500}), respectively.}
\label{nmc}}
\end{center}
 \end{figure}

Unfortunately the data on electroproduction of charmonia are
very scanty so far. There are only data from the NMC experiment 
\cite{NMC-psi} concerning energy dependence of the ratio of nuclear 
transparencies $Tr_{Sn}^{inc}$ and $Tr_{C}^{inc}$ for
incoherent production of $J/\Psi$ at $Q^2 = 0$.
The corresponding photon energy varies from 60 to 210 GeV.
It allows to study the transition from medium long to long
coherence length, which varies from 2.4 to 8.5$\fm$.
For long $l_c\gsim 8.5\fm$
the ``frozen'' approximation can be used with high accuracy.
In this case nuclear transparency $Tr_A^{inc}$  of incoherent 
(quasielastic) $J/\Psi$ production can be calculated  
using Eq.~(\ref{520}) and the simplified ``frozen''
approximation Eqs.~(\ref{465}) -- (\ref{510}). 
For medium long coherence length one can not use the ``frozen''
approximation and fluctuations of the size of the $\bar qq$ pair become  
important.
Because of a strong inequality $l_c < l_f$ for charmonium production
CT effects are expected to be dominant at small and moderate energies.
Consequently, 
they should lead to a rise with energy
of $Tr_{A}^{inc}$.
Such a scenario is depicted in Fig.~\ref{nmc} 
by solid and dashed curves.
Dashed curve show our results using the LC Green function
approach in the limit of short coherence length
$l_c\to 0$  Eq.~(\ref{500}).
The solid one includes in addition also CL effects. 
Thus, the effect of coherence length manifest itself as 
a separation between the solid and dashed curves.
A rise with energy of the ratio
$Tr_{Sn}^{inc}/Tr_{C}^{inc}$ at small and medium energy is a net
manifestation of CT. 
It follows from a rise of formation time, see Eq.~(\ref{20}).
At larger energies when CL effects become also important
the ratio $Tr_{Sn}^{inc}/Tr_{C}^{inc}$ starts to be 
gradually smaller\footnote{
In energy dependence of nuclear transparency at fixed $Q^2$
effect of the coherence
follows from variation of the coherence length from small to 
large values compared to the nuclear size, see Eq.~(\ref{30})}.
Unfortunately, the NMC data have quite large error bars
and therefore give only an indication for such a behavior.
More accurate data are needed for exploratory study of
CT and CL effects. Charmonium real photoproduction off nuclei
at small and large energies 
is very sensitive for investigation of CT and CL effect separately.
Not so for real photoproduction of light vector mesons
when coherence and formation lengths are comparable
and mixing of CT and CL effects exists already at small energies.

Problem of separation of CT and CL effects was 
discussed in details
in \cite{knst-01} with the main emphasis to production   
of light vector mesons where $l_c \gsim l_f$ at $Q^2 \lsim 1\div 2\GeV^2$.
In this paper we present the results for charmonium production
where a strong inequality $l_c < l_f$ in all discussed kinematic region 
leads to a different scenario of CT-CL mixing as compared 
with light vector mesons.
Consequently, at fixed $Q^2$ and at small and medium energies 
the problem of CT-CL separation is not so arisen.   
In order to eliminate the effect of CL from the data 
on the $Q^2$ dependence of nuclear transparency one should bin the data in a 
such way which keeps $l_c = const$ \cite{hk-97}. 
It means that one should vary simultaneously $\nu$
and $Q^2$ maintaining the CL Eq.~(\ref{30}) constant,
 \beq
\nu = {1\over2}\,l_c\,(Q^2+m_{J/\Psi}^2)\ .
\label{534}
 \eeq
 In this case the Glauber model predicts a $Q^2$ independent nuclear
transparency, and any rise with $Q^2$ would signal CT \cite{hk-97}.

\begin{figure}[tbh] 
\includegraphics{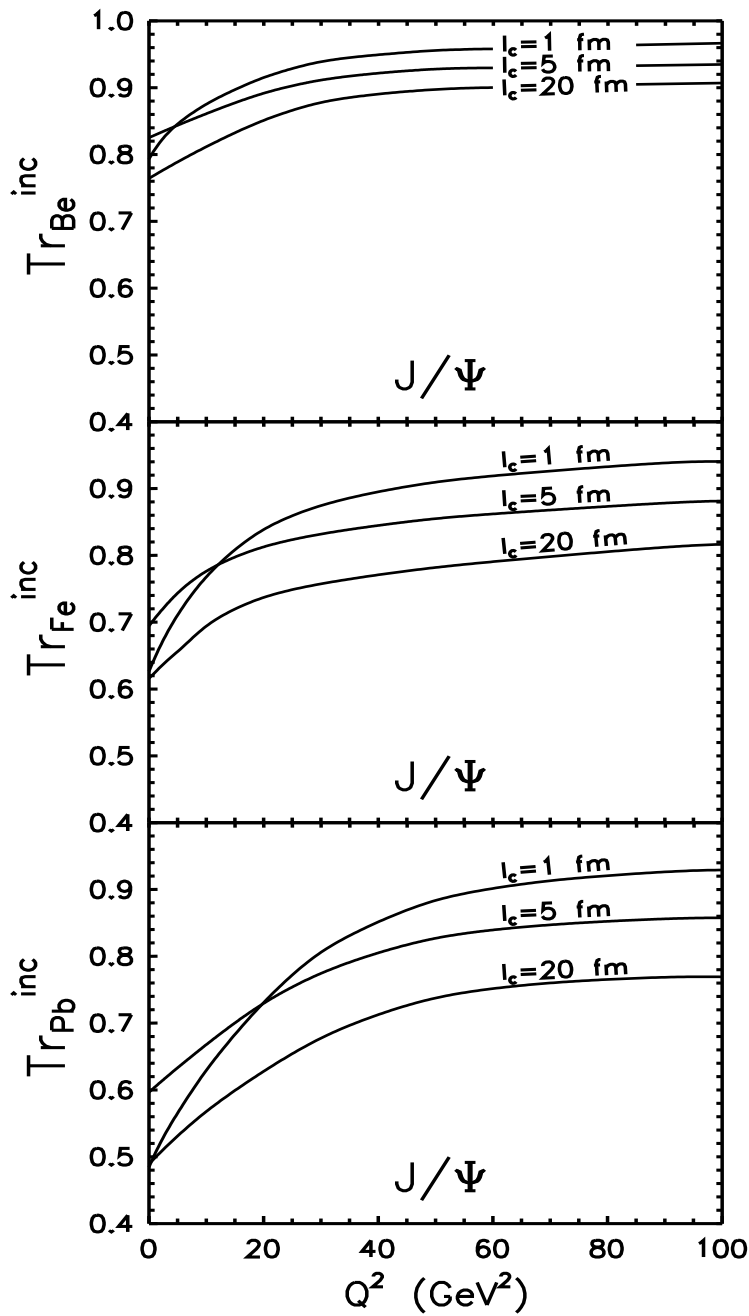} 
\begin{center}
\vspace{17cm} 
\parbox{13cm} 
{\caption[Delta] 
 {$Q^2$ dependence of the nuclear transparency $Tr_A^{inc}$ for
exclusive electroproduction of $J/\Psi$  
on nuclear targets $^{9}Be$, $^{56}Fe$ and $^{207}Pb$ (from top to
bottom). The CL is fixed at $l_c = 1$, $5$ and
$20\fm$.}
 \label{lc-const-inc}}
\end{center}
 \end{figure}

The LC Green function technique incorporates both the effects of
coherence and formation. We performed calculations of $Tr_A^{inc}(Q^2)$
at fixed $l_c$ starting from different minimal values of $\nu$, which
correspond to real photoproduction in Eq.~(\ref{534}),
 \beq
\nu_{min}={1\over2}\,l_c\,m_{J/\Psi}^2\ . 
\label{536} 
 \eeq
The results for incoherent production of $J/\Psi$ at 
$\nu_{min}= 24.3,\ 121.7$ and $487\GeV$ ($l_c=1, 5$ and 
$20\fm$) are presented in Fig.~\ref{lc-const-inc} for beryllium, 
iron and lead. 
 We use the nonperturbative LC wave function of the photon with the
parameters of the LC potential $a_{0,1}$ fixed in accordance with
Eq.~(\ref{190}) at $v=1/2$. We use quark mass $m_c=1.5\GeV$. 

Although the predicted variation of nuclear transparency with
$Q^2$ at fixed $l_c$ is less than for production of light vector
mesons \cite{knst-01} it is still rather significant
to be investigated experimentally even in the range of $Q^2\lsim
20\GeV^2$. 
CT effects (the rise with $Q^2$ of nuclear transparency)
are more pronounced at low than at high energies
and can be easily identified by the planned future
experiments.

\begin{figure}[htb]
\includegraphics{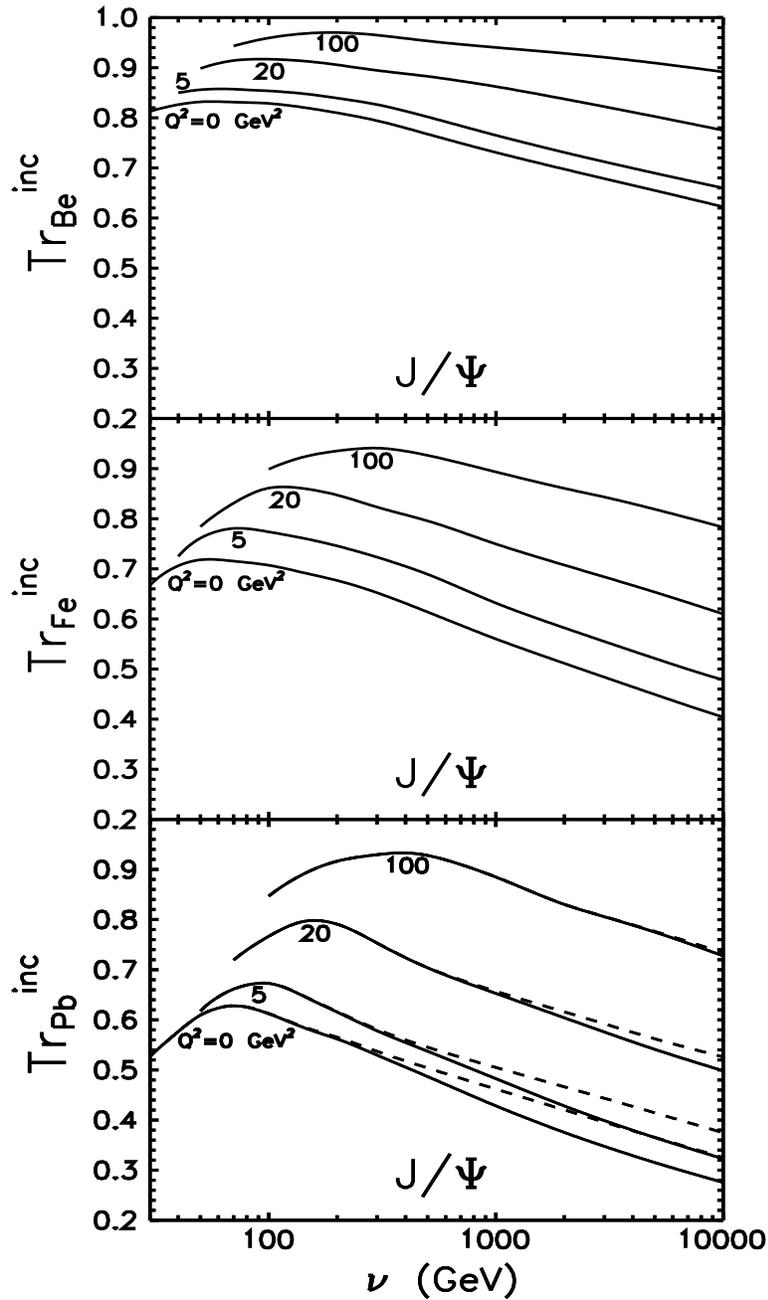}
\begin{center}  
\vspace{17cm}
\parbox{13cm}
{\caption[Delta]
 {Nuclear transparency for incoherent electroproduction $\gamma^*A\to
J/\Psi~X$ as function of energy at $Q^2=0,\ 5,\ 20$ and $100\GeV^2$ for
beryllium, iron and lead. The solid curves and dashed curves for lead 
correspond to calculations with and without gluon shadowing,
respectively.}
 \label{e-incoh}}
\end{center}
 \end{figure}

We also calculated the energy dependence of nuclear transparency at fixed 
$Q^2$. The results for beryllium, iron and lead are shown  in 
Fig.~\ref{e-incoh} for different values of $Q^2$.
 The interesting feature is the presence of a maximum of transparency at
some energy which is much more evident than in production of light
vector mesons \cite{knst-01}.
At small and moderate energies a strong rise of $Tr_{A}^{inc}$ with energy
is a manifestation of net CT effects as a result of a strong
inequality $l_c < l_f$. 
The existence of maxima of $Tr_{A}^{inc}$
results from the interplay of coherence and formation
effects. Indeed, the formation length (FL) 
rises with energy leading to an increasing
nuclear transparency. At some energy, however, the effect of CL which is
shorter than the FL, is switched on leading to a growth of the path
length of the $\bar qq$ in the nucleus, i.e. to a suppression of
nuclear transparency. 
This also explains the unusual ordering of curves calculated for different 
values of $l_c$ as is depicted in Fig.~\ref{lc-const-inc}.

\section{Coherent production of $J/\Psi$}
\label{psi-coh}

First of all we present a short introduction to coherent production
of vector mesons.
One should replace $V\rightarrow J/\Psi$ and $\bar qq\rightarrow\bar cc$
when coherent production of charmonia is treated.
In general,
in coherent (elastic) electroproduction of a vector mesons 
the target nucleus remains intact, so
all the vector mesons produced at
different longitudinal coordinates and impact parameters add up
coherently. This condition considerably simplifies the expressions 
for the production cross sections. The
integrated cross section has the form,
 \BE
\sigma_A^{coh}\equiv
\sigma_{\gamma^{*}A\to VA}^{coh} = 
\int d^2q\,\left|\int d^2b\,
e^{i\vec q\cdot\vec b}\,
{\cal M}_{\gamma^{*}A\to VA}^{coh}(b)
\right|^2 = 
\int d^{2}\,{b}\,
|{\cal M}_{\gamma^{*}A\to VA}^{coh}
({b})\,|^{2}\ ,
\label{550}
 \EE
 where
 \BE
{\cal M}_{\gamma^{*}A\to VA}^{coh}({b}) =
\int\limits_{-\infty}^{\infty}\,dz\,\rho_{A}({b},z)\,
F_{1}({b},z)\ ,
\label{560}
 \EE
 with the function $F_{1}({b},z)$ defined in (\ref{505}).

One should not use Eq.~(\ref{485}) for nuclear transparency any more
since the $t$-slopes of the differential cross sections for nucleon and
nuclear targets are different and do not cancel in the ratio. Therefore,
the nuclear transparency also includes the slope parameter
$B_{\gamma^*N}$ for the process $\gamma^{*}\,N\rightarrow V\,N$,
 \BE
Tr_{A}^{coh} = \frac{\sigma_{A}^{coh}}{A\,\sigma_{N}} = 
\frac{16\,\pi\,B_{\gamma^*N}\,\sigma_{A}^{coh}}{A\,
|{\cal M}_{\gamma^{*}N\to VN}(s,Q^{2})\,|^{2}}
\label{570}
 \EE

The energy dependent factor $C(s)$ 
in dipole cross section approximation 
Eq.~(\ref{460}) is adjusted in the limit
$l_c\gg R_A$ 
in  analogical way as
for incoherent production of charmonia 
described in the previous section.
However, in contrast to Eq.~(\ref{467})
the factor $C(s)$ is fixed now 
by the following relation,
 \beqn
&& \frac{
\int d^{2}b\,\left|\int d^{2}r\,\int d\alpha\,
\Psi_{V}^{*}(\vec r,\alpha)\,
\Psi^{T,L}_{\gamma^{*}}(\vec r,\alpha)\,
\left\{1 - \exp\biggl[ - {1\over2}\, 
C_{T,L}(s)\,r^2\,T_{A}({b})\,\biggr]
\right\}\right|^2 }
{\left|\int d^{2}r\,\int d\alpha\,
\Psi_{V}^{*}(\vec r,\alpha)\,C_{T,L}(s)\,r^2\,
\Psi^{T,L}_{\gamma^{*}}(\vec r,\alpha)\right|^2}  
\nonumber\\ &=&
\frac{
\int d^{2}{b}\,\left|\int d^{2}r\,\int d\alpha\,
\Psi_{V}^{*}(\vec r,\alpha)\,
\Psi^{T,L}_{\gamma^{*}}(\vec r,\alpha)\,
\left\{1 - \exp\biggl[ - {1\over2}\, 
\sigma_{\bar qq}(r,s)\,T_{A}({b})\biggr]
\right\}\right|^2}{\left|\int d^{2}r\,\int d\alpha\,
\Psi_{V}^{*}(\vec r,\alpha)\,\sigma_{\bar qq}(r,s)\,
\Psi^{T,L}_{\gamma^{*}}(\vec r,\alpha)\right|^2 }\, .
\label{468}
 \eeqn

\subsection{Predictions for coherent production of $J/\Psi$}
\label{coh-data}

Unfortunately, there are no data yet on coherent electroproduction
of charmonia therefore we present only predictions which can be
later verify and tested in the future planned experiments.

 \begin{figure}[tbh] 
\includegraphics{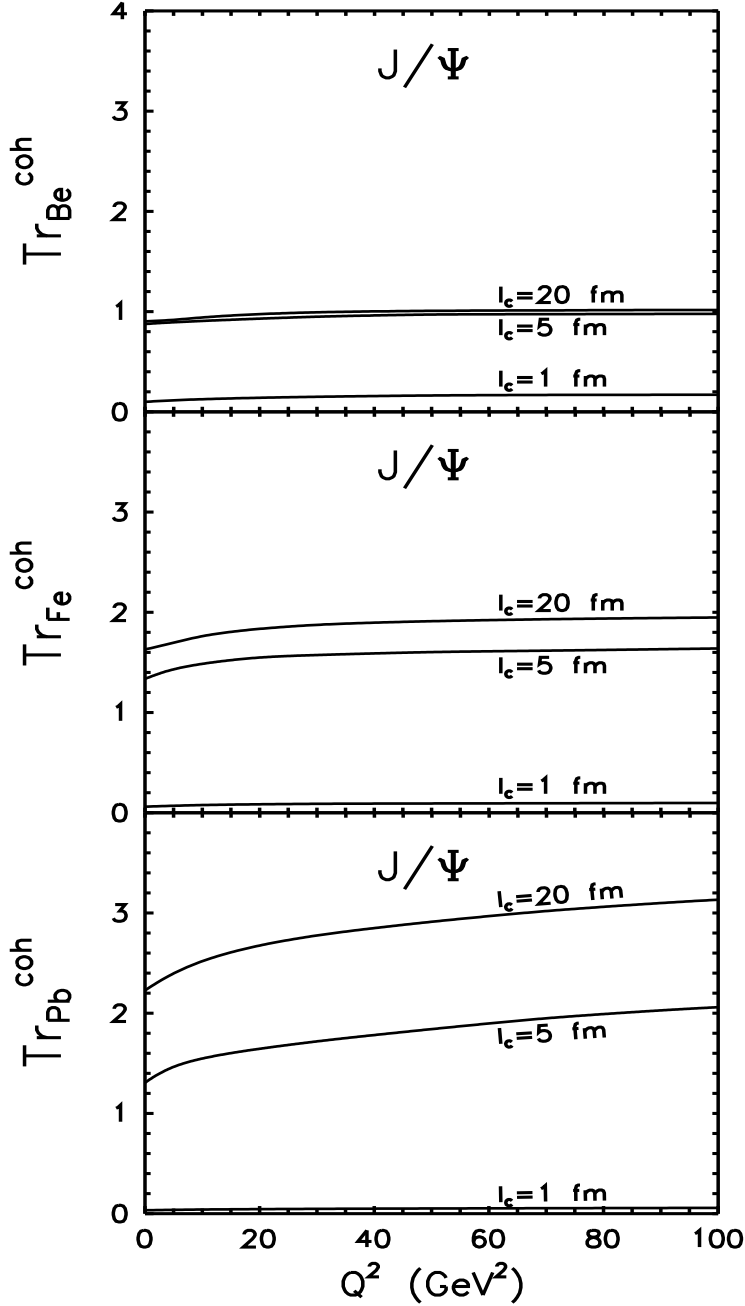} 
\begin{center}
\vspace{17.0cm} 
\parbox{13.0cm} 
{\caption[Delta] 
 {The same as in Fig.~\ref{lc-const-inc}, but for coherent production
of $J/\Psi$, $\gamma^*A\to J/\Psi~A$.}
 \label{lc-const-coh}}
\end{center}
 \end{figure}

One can eliminate the effects of CL and single out the net CT effect in a
way similar to what was suggested for incoherent reactions by selecting
experimental events with $l_c=const$. We calculated nuclear transparency
for the coherent reaction $\gamma^*A\to J/\Psi A$ at fixed values of
$l_c$. The results for $l_c=1,\ 5$ and $20\fm$ are depicted in
Fig.~\ref{lc-const-coh} for several nuclei.
We performed calculations of $Tr_A^{coh}$ with the slope $B = 4.7\GeV^{-2}$.
The effect of a rise of $Tr_{A}^{coh}$ is not sufficiently large to be 
observable in the range $Q^2\leq 20\GeV^2$.
A wider range $Q^2\leq 100\GeV^2$ and heavy nuclei gives
higher chances for experimental investigation of CT. However,
on the other hand, it encounters the problem of low yields at high $Q^2$.

Note that in contrast to incoherent production where nuclear transparency
is expected to saturate as $Tr^{inc}_A(Q^2) \to 1$ at large $Q^2$, for
the coherent process nuclear transparency reaches a higher limit,
$Tr^{coh}_A(Q^2) \to A^{1/3}$.

 \begin{figure}[htb]
\includegraphics{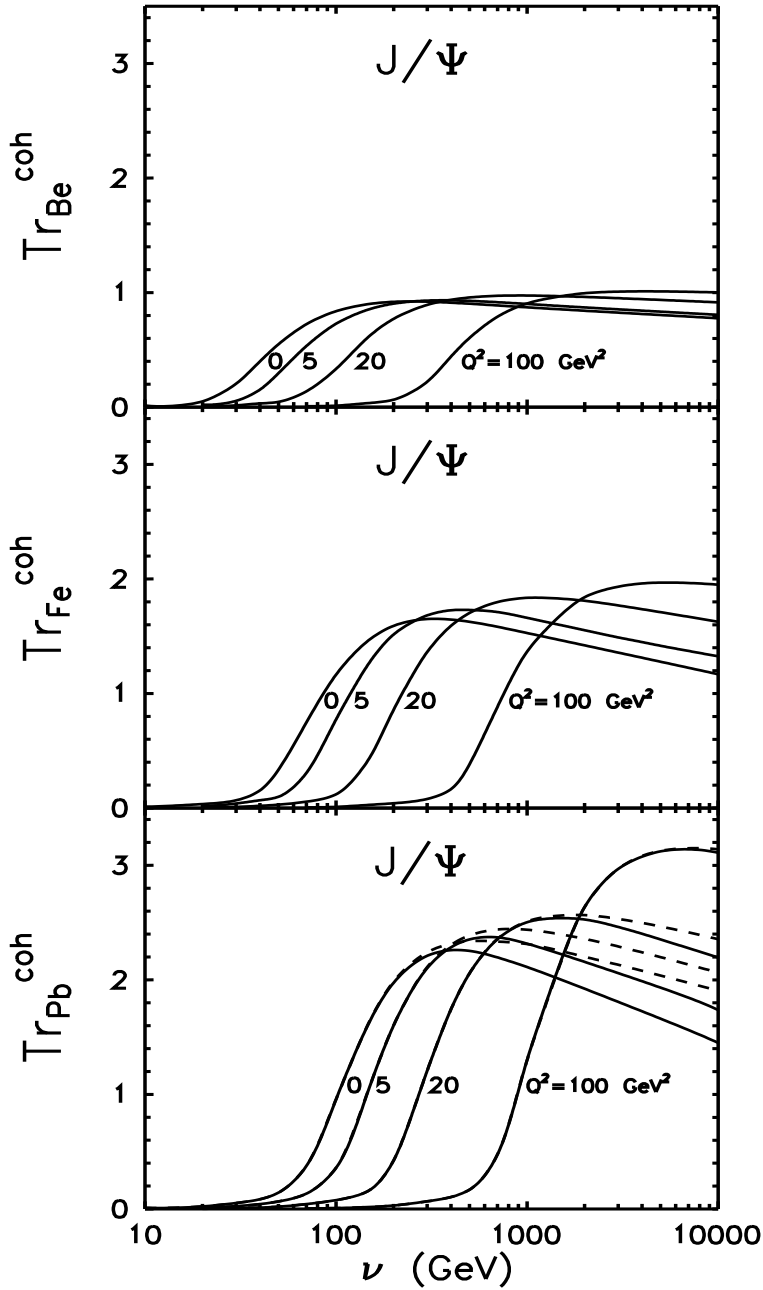}
\begin{center}
\vspace{17.0cm}
\parbox{13cm}
{\caption[Delta]
 {Nuclear transparency for coherent electroproduction 
$\gamma^*A\to J/\Psi~A$ as function of energy at $Q^2=0,\ 5,\
20$ and $100\GeV^2$
for beryllium, iron and lead.
The solid curves and dashed curves for lead 
correspond to calculations with and without gluon shadowing,
respectively.}
 \label{e-coh}}
\end{center}
 \end{figure}

We also  calculated nuclear transparency as function of energy at fixed 
$Q^2$.
The results for $J/\Psi$ produced coherently off beryllium, iron
and lead 
are depicted 
in Fig.~\ref{e-coh} at $Q^2=0,\ 5,\ 20$ and $100\GeV^2$. 
$Tr^{coh}_A$ is very small at low energy, what of course does not mean
that nuclear matter is not transparent, but the nuclear coherent cross
section is suppressed by the nuclear form factor. Indeed, the
longitudinal momentum transfer which is equal to the inverse CL, is large
when the CL is short. However, at high energy $l_c\gg R_A$ and nuclear
transparency nearly saturates (it decreases with $\nu$ only due to the
rising dipole cross section). The saturation level is higher at larger
$Q^2$ which is a manifestation of CT.

Note that in all the calculations the effects of gluon shadowing are 
included
in an analogical way as 
described in detail in the recent papers
\cite{knst-01,ikth-02}.
They are much smaller than in production of light vector mesons.
For illustration they are depicted in Figs.~\ref{e-incoh} and
\ref{e-coh} for the lead target 
as a difference between solid and dashed lines 
at various values of $Q^2$.
In the photoproduction limit $Q^2 = 0$ 
the onset of gluon shadowing becomes
important at rather high photon energy $\nu > 1000\GeV$
for incoherent and $\nu > 500\,\GeV$ for coherent
production. This corresponds to the
claim made in \cite{kst2} that the onset of gluon shadowing requires
smaller $x_{Bj}$ than the onset of quark shadowing. The reason is 
that the fluctuations containing gluons 
are in general heavier than the $\bar qq$
and have a shorter CL.

Although gluon shadowing is included in all calculations,
in the kinematic range 
discussed in the present paper
it is small enough 
and consequently does not affect the main achievements
and conclusions important
for investigation of CT effects in coherent and
incoherent production of charmonia off nuclei.

\section{Summary and conclusions}
\label{conclusions}

In the present paper we focused the main emphasis 
to production of charmonia due to
advantages as compared with production of light vector mesons
\cite{knst-01}.
Electroproduction of charmonia off nuclei is very convenient 
for study of an interplay
between coherence (shadowing) and formation (color transparency) effects.
A strong inequality $l_c < l_f$ in all kinematic region
of $\nu$ and $Q^2$
leads to a different scenario of mixing of CT and CL effects
as compared with light vector mesons
where $l_c \gsim l_f$ at $Q^2\lsim 1\div 2\GeV^2$.
Consequently, 
at small and moderate energies
a problem of CT-CL separation is not so arisen.
Besides, due to quite a large mass $m_c = 1.5\GeV$ the 
relativistic corrections and nonperturbative effects are sufficiently smaller.
Investigation of still heavier vector mesons (bottonium, toponium)
encounters the problem of very low yields as well as very small CT and CL
effects (due to very large mass of $\bar qq$ fluctuactions and of vector
meson) to be 
measured experimentally.
Therefore production of charmonia represent some compromise because 
above mentioned theoretical uncertainties typical for 
light vector mesons and very small production rates
typical for still heavier vector mesons are eliminated to a certain
extent keeping
sufficiently large CT and CL effects to be detected experimentally. 
This fact supports an enhanced interest to study 
electroproduction of charmonia off nuclei separately.
We used from \cite{knst-01} a rigorous
quantum-mechanical approach based on the light-cone QCD Green function
formalism which naturally incorporates the interference 
effects of CT and CL.
Our main results and observations are the following.

\begin{itemize}

\item 
Within suggested approach taken from \cite{knst-01} 
interpolating between the previously known low and
high energy limits
we studied for the first time CT effects in incoherent and coherent 
electroproduction of charmonia off nuclei.

\item 
As the first test we compare the model predictions with available
data from the NMC experiment concerning energy dependence
of the ratio $Tr_{Sn}^{inc}/Tr_{C}^{inc}$ 
of nuclear transparencies for incoherent production of $J/\Psi$
at $Q^2 = 0$. We found a good agreement with the data what
confirms a dominance of CT effects at small and medium 
and CL effects at medium large
and large energies. 

\item 
The onset of coherence effects (shadowing) can mimic the expected
signal of CT in incoherent electroproduction of charmonia at medium large
and large energies. In order to single out the formation effect data must
be taken at such energy and $Q^2$ which keeps $l_c = const$. 
Then observation of a rise with $Q^2$ of nuclear
transparency for fixed $l_c$ would give a signal of color
transparency.
Predictions of $Tr_A^{inc}(Q^2)$ as a function of $Q^2$
at different fixed $l_c$ shows rather large CT effects
in incoherent production of charmonia.
Although the variation with $Q^2$ of nuclear transparency 
at fixed $l_c$ is predicted to be less 
than for production of light vector
mesons \cite{knst-01} it is still rather significant
to be investigated experimentally even in the range $Q^2\lsim
20\GeV^2$.
CT effects (the rise with $Q^2$ of nuclear transparency)
are more pronounced at low than at high energies
and can be easily identified by the planned future
experiments.  

\item 
The effects of CT in coherent production of charmonia are found
to be less pronounced similarly as in production of light vector mesons
\cite{knst-01}.
A wider range $Q^2\leq 100\GeV^2$ and heavy nuclei gives
higher chances for experimental investigation of CT. However,
on the other hand, it faces the problem of low yields at high $Q^2$.

\item 
The effects of gluon shadowing were shown to be important only 
at much higher energies than in production of light vector mesons
due to large mass of $\bar cc$ fluctuation.
Nuclear suppression of gluons was calculated
within the same LC approach and included in predictions.
It was manifested that these corrections are quite a small 
at medium energies which are dominant
in the process of searching for CT effects.

\item 
Finally one can
compare the predictions of charmonium incoherent and coherent production off 
lead target
(see Figs.~\ref{lc-const-inc}, ~\ref{e-incoh}, ~\ref{lc-const-coh} and
\ref{e-coh}) obtained within rigorous
quantum-mechanical approach based on the
light-cone QCD Green function formalism (incorporating naturally
CT and CL effects)
with
results of the paper \cite{ikth-02} evaluated in the approximation of
long coherence length $l_c \gg R_A$ (without CT effects) 
with realistic light-cone wave
functions of charmonia and
with corrections for finite values of $l_c$.
We find a nice quantitative agreement at moderate and high energies
and at low and medium values of $Q^2$.
This fact confirms justification to use that
high-energy approximation \cite{ikth-02}
for charmonium electroproduction off nuclei in the kinematic
region where CL effects dominate. 
Moreover, it gives a basis to perform in the future the fully realistic 
calculations
within LC dipole approach based on the light-cone Green function formalism
employing a realistic dipole cross section 
and realistic LC wave functions of charmonia from \cite{hikt-00}. 

\end{itemize}

Concluding, the predicted rather large effects of CT in incoherent
electroproduction of charmonia off nuclei open further possibilities for
the search for CT with medium energy electrons and can be tested
in future experiments.

\medskip

\noindent
 {\bf Acknowledgments}:
This work has
been supported in part by the Slovak Funding Agency, Grant No. 2/1169 and
Grant No. 6114.

\end{document}